\documentclass[preprint,showpacs,preprintnumbers,amsmath,amssymb, showkeys,superscriptaddress,titlepage,8pt]{revtex4-2}
\usepackage{graphicx}% Include figure files
\usepackage{dcolumn}% Align table columns on decimal point
\usepackage{bm}% bold math
\usepackage{hyperref}
\usepackage{textcomp}

\begin{document}

\title{BRIEF REVIEW ON UNSTABLE STATES IN DISSOCIATION OF RELATIVISTIC NUCLEI}

\author{A.A. Zaitsev}
\email{zaicev@jinr.ru}
\affiliation{Joint Institute for Nuclear Research (JINR), Dubna, Russia}

\begin{abstract}
The article presents a review of studying unstable states in the relativistic dissociation of nuclei $^{10}$B, $^{11,12}$C, $^{16}$O, $^{22}$Ne, $^{28}$Si, $^{84}$Kr and $^{197}$Au in the energy range from hundreds of MeV/nucleon to several tens of GeV/nucleon using the nuclear track emulsion method. A systematic study of the fragmentation of incident nuclei with multiple formation of the lightest fragments of He and H. This made it possible to study the dynamics of unstable nuclear states of $^8$Be, the Hoyle state and the 4$\alpha$-particle state of the $^{16}$O nucleus above the threshold in the relativistic dissociation of nuclei due to precision measurements of fragment emission angles. It is shown that to reconstruct the relativistic decays of unstable nuclei in a nuclear track emulsion, it is sufficient to determine the invariant mass of the system of He and H fragments in the approximation of conservation of momentum per nucleon of the parent nucleus. This approach allows one to search for more complex nuclear states. An indication has been obtained of growing probability to detect $^8$Be with increasing the number of relativistic $\alpha$ particles in the event.
\end{abstract}

\maketitle 

\section{Introduction}

One of the key aspects of the nuclear structure is the presence of degrees of freedom where quartets of spin-paired protons and neutrons behave as constituent clusters. It is manifested in the intense production of $\alpha$ particles in a wide variety of nuclear reactions and decays. The transition to the study of ensembles of $\alpha$ particles immediately above the bound thresholds allows one to reveal the role of unstable $^8$Be and $^9$B nuclei and the 3$\alpha$ Hoyle state (HS) and to search for their analogues.

The use of a technically simple and inexpensive method of nuclear track emulsion (NTE) in beams of relativistic nuclei provides flexibility and uniformity at the search stage, and in the theoretical aspect - transparency of interpretation. It is very important to demonstrate the similarity of conclusions based on relativistic invariance. During the dissociation of relativistic nuclei in a narrow solid angle of fragmentation, the ensembles of He and H nuclei are intensively generated (Fig. 1). Particularly important are the so-called “white” stars in the region of the interaction vertex where no tracks of the target nucleus and produced mesons are observed. In besides, when tracking in the direction of the fragmentation cone of heavy nuclei, one can observe stars which do not have an incoming track from the top of the event. These stars appear in the interactions of relativistic neutron fragments and nuclei from the NTE composition [1].

According to the widths, the decays of $^8$Be, $^9$B and HS occur over ranges from several thousands ($^8$Be and HS) to several tens ($^9$B) of atomic sizes and should be identified by a minimum invariant mass. Due to the minimum energy, the decays of
 $^8$Be, $^9$B and HS should appear as pairs and triplets with relativistic He and H fragments having the smallest expansion angles. The invariant mass of a system of relativistic fragments is defined as the 
 sum of all scalar products of 4-momenta $P_{i,k}$ fragments $M^{*2} = \Sigma(P_i\cdot P_k)$. For convenience of presentation, the variable $Q$ is adopted and defined as the difference between the invariant 
 mass and the sum of the 
 fragment masses $Q = M^* - \Sigma m$. 
 The components $P_{i,k}$ are determined in the approximation of conservation of initial momentum per nucleon by fragments.

Contemporary interest in nuclear $\alpha$-clustering is largely motivated by the concept of $\alpha$-particle Bose–Einstein condensate ($\alpha$BEC). The unstable $^8$Be and HS nuclei are described as 2- and 3$\alpha$BEC states. Their decays can serve as signatures of the decays of more complex $n\alpha$BEC states. The existence of the latter can expand the picture of the nucleosynthesis of heavy nuclei. Recently, the statistics of dozens of $^8$Be decays has revealed the growing probability to detect $^8$Be with increasing the number of associated $\alpha$ particles ($n\alpha$). A preliminary conclusion has been made that the contributions from the $^9$B and HS decays are also growing.

\begin{figure}[h]
	\center{\includegraphics[width=1\linewidth]{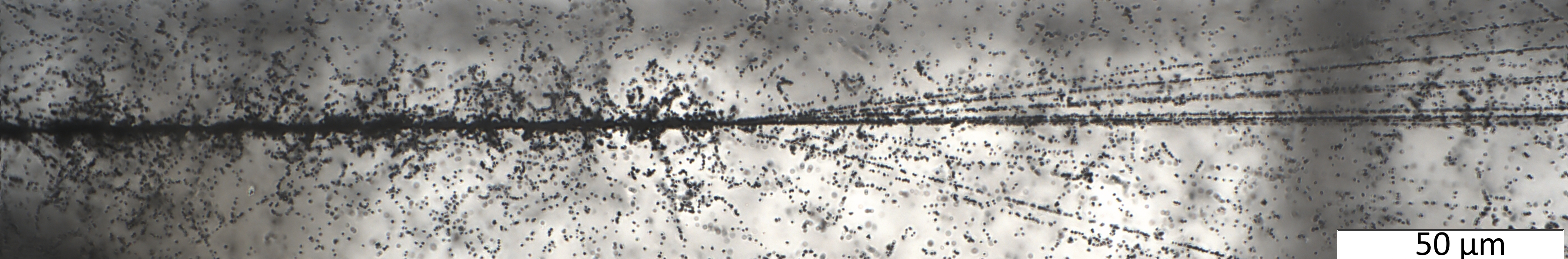}}
	\caption{Macrophoto of the peripheral interaction of the $^{84}$Kr nucleus with the energy of 950 MeV per nucleon having multiple formation of secondary He and H fragments.}
	\label{fig1}
\end{figure}

\section{IDENTIFICATION OF DECAYS OF $^8$Be AND $^9$B NUCLEI IN DISSOCIATION OF LIGHT RELATIVISTIC NUCLEI}

The analysis of the exposure of NTE layers in the beam of $^{10}$B nuclei at the energy of 1 $A$ GeV made it possible to reveal the effect of dominance of the multiple fragmentation channel [2, 3]. In the distribution of the fragments by a charge state, the share of the $^{10}$B $\to$ 2He + H channel was 77\%. Based on measurements of the emission angles of the He and H fragments, it has been established that the unstable nucleus $^8$Be$_{g.s.}$ (Fig. 2, right) appears with the probability of (25 $\pm$ 5)\% where (13 $\pm$ 3)\% take place due to the decays of the unstable $^9$B nucleus (Fig. 2, left). What has been unexpected is the fact that the number of ``white'' stars $^9$B + n is 10 times bigger than $^9$Be + p. This observation may indicate a wider spatial distribution of neutrons in the $^{10}$B nucleus compared to the protons, which results in a larger cross section for the $^9$B+n channel compared to the mirror channel. In addition, with the probability of 8\% the stars have been observed in the $^{10}$B $\to$ $^6$Li + $\alpha$ channel. It is possible that the Li nucleus, weakly manifesting in the dissociation of $^{10}$B, is also present in $^{10}$B mainly in a ``dissolved'' form making a non-resonant contribution to the $\Theta_{2He}$ distribution [2-4].

\begin{figure}[h]
	\center{\includegraphics[width=1\linewidth]{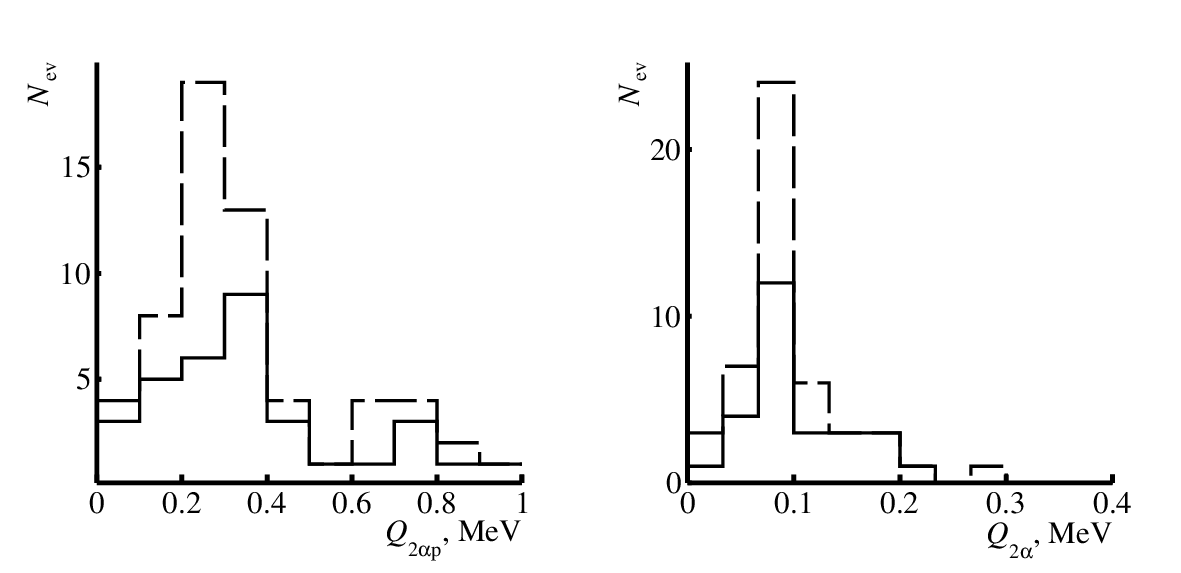}}
	\caption{Distributions of 2$\alpha$p triplets by excitation energy $Q_{2\alpha p}$ (left figure) for fragmentation $^{10}$B $\to$ 2He + H at 1.6 $A$ GeV/$c$ (solid line) and $^{11}$C $\to$ 2He + 2H at 2.0 $A$ GeV/$c$ (added, dotted line) and $Q_{2\alpha}$ of $\alpha$ pairs in $^9$B decays identified in these events (right figure) [7].}
	\label{fig2}
\end{figure}

The charge topology of the dissociation channels of $^{11}$C nuclei in NTE with the energy of 1.2 $A$ GeV has been studied. Among $^{11}$C stars the events with observations of only relativistic fragments of He and H, especially 2He + 2H, predominate: their contribution was 77\% [5]. The channel characteristic only of the $^{11}$C nucleus with the formation of Li+He+H fragments has been established. Based on the measured emission angles of He and H fragments, presented in the invariant variable $Q$, it has been shown that decays of $^8$Be$_{g.s.}$ nuclei (Fig. 2, right) of all the found ``white'' $^{11}$C stars are shown in 21\% of $^{11}$C$\to$2He + 2H events and 19\% - for the $^{11}$C$\to$3He channel. The $^9$B decays (Fig. 2, left) have been detected in the ``white'' stars $^{11}$C $\to$ 2He + 2H, constituting 14\% of the ``white'' stars $^{11}$C $\to$ 2He + 2H. It has been found that, as in the case of $^{10}$C, $^8$Be$_{g.s.}$ the decays of the ``white'' $^{11}$C stars almost always appear due to the decays of $^9$B [5,6]. It is worth noting that the lowest energy peak in the $Q_{2\alpha2p}$ distribution of 18 found stars $^{11}$C $\to$ 2He + 2H are characterized by the average $Q_{2\alpha2p}$ value of (2.7 $\pm$ 0.4) MeV with the RMS value of 2.0 MeV [6].

The limitation established during the analysis of the data on the dissociation of $^{10}$B and $^{11}$C nuclei while identifying the decays of $^8$Be nuclei ($Q_{2\alpha}$ $<$ 0.2 MeV) has allowed one to estimate the contribution of these decays to the dissociation of relativistic nuclei $^{12}$C$\to$ 3$\alpha$ and $^{16}$O$\to$ 4$\alpha$ in NTE at the level of (45 $\pm$ 4)\% and ( 62 $\pm$ 3)\%, respectively (Fig. 3) [7-9].

\begin{figure}[h]
	\center{\includegraphics[width=0.65\linewidth]{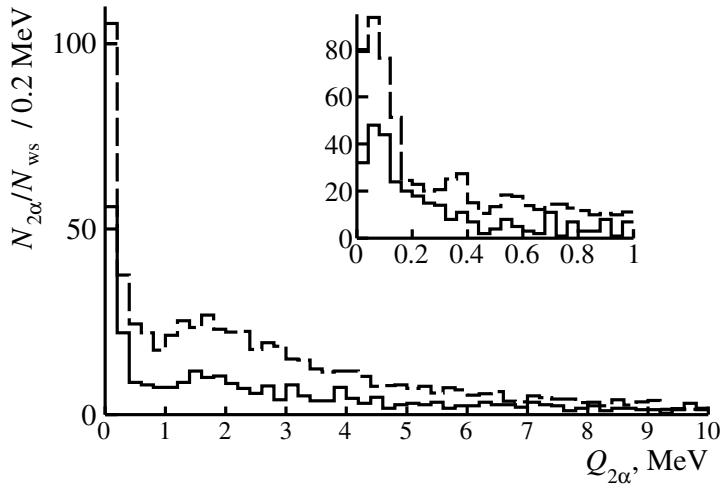}}
	\caption{Distribution of the number of 2$\alpha$-pairs $N_{2\alpha}$ according to the excitation energy $Q_{2\alpha}$ in the coherent dissociation $^{12}$C $\to$ 3$\alpha$ (solid line) and $^{16}$O $\to$ 4$\alpha$ (dashed line) at 3.65 $A$ GeV; inset: enlarged region $Q_{2\alpha}$ $<$ 1 MeV (step 40 keV); histograms have been normalized to the number of ``white'' stars $N_{ws}$ [9].}
	\label{fig3}
\end{figure}

\section{OBSERVATION OF EVENTS WITH THE HOYLE STATE DECAYS}

The certainty in the identification of $^8$Be and $^9$B has become the ground to search for HS decays in the $^{12}$C $\to$ 3$\alpha$ dissociation (Fig. 4), where the limit on the $Q$ variable of $\alpha$-triples was set to 0.7 MeV [7]. A comprehensive analysis of $^{12}$C $\to$ 3$\alpha$ and $^{16}$O $\to$ 4$\alpha$ stars has made it possible to determine that the fraction of events containing HS decays is (11 $\pm$ 3)\% for $^{12}$C and (22 $\pm$ 2)\% - for $^{16}$O (Fig. 4) [7-11]. The 33 $^{16}$O $\to$ 2$^8$Be events have been identified, accounting for 5 $\pm$ 1\% of the ``white'' $^{16}$O $\to$ 4$\alpha$ stars. The distribution of the $Q$ value of a system of 4$\alpha$ particles in $^{16}$O $\to$ 2$^8$Be events [12] has found two candidates $^{16}$O(0$^+_6$) $\to$ 2$^8$Be in the region $Q <$ 1.0 MeV. The dissociation statistics for the $^{16}$O $\to$ 2$^8$Be and $^{16}$O $\to$ $\alpha$HS channels has the following ratio: 0.22 $\pm$ 0.02 [9].

\begin{figure}[h]
	\center{\includegraphics[width=0.65\linewidth]{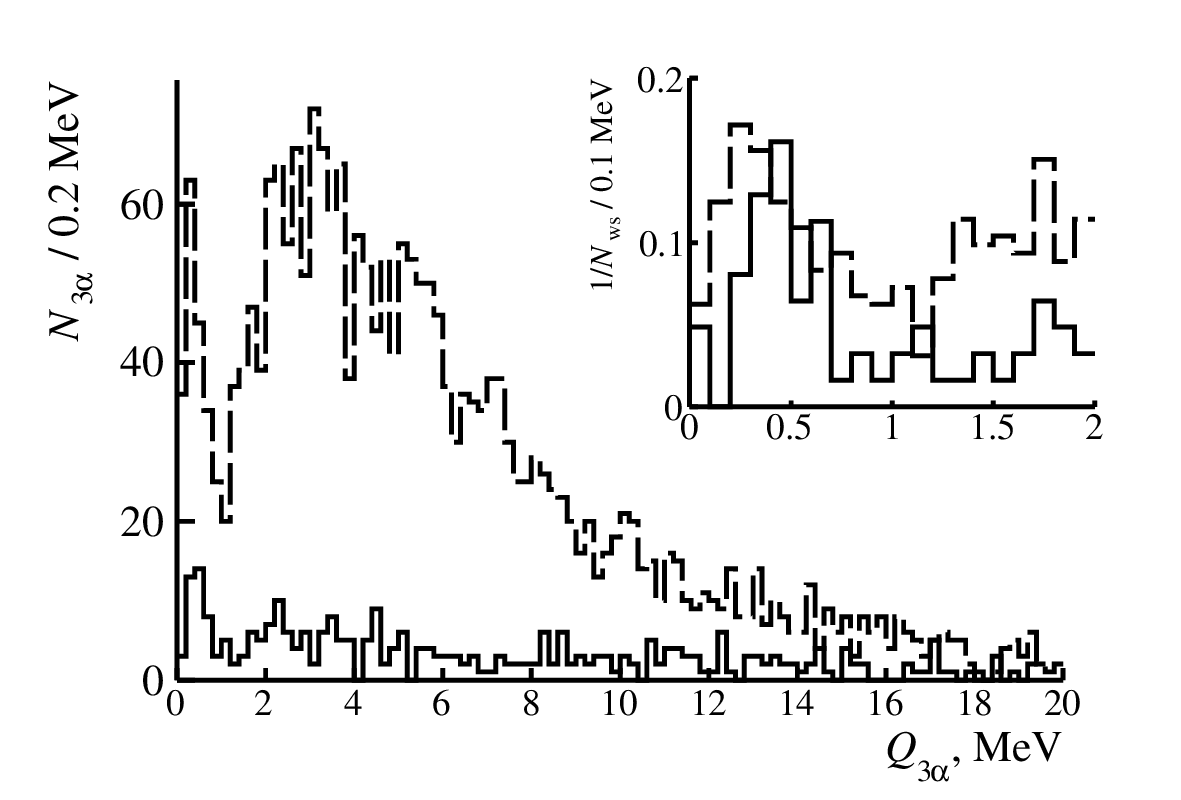}}
	\caption{Distribution of the number of 3$\alpha$-triples $N_{3\alpha}$ by excitation energy $Q_{3\alpha}$ for 316 ``white'' stars $^{12}$C $\to$ 3$\alpha$ (solid) and 641 ``white'' stars $^{16}$O $\to$ 4$\alpha$ (dashed) at the energy of 3.65 $A$ GeV; The inset shows the enlarged part of $Q_{3\alpha}$ $<$ 2 MeV, normalized to the number of ``white'' stars $N_{ws}$ [9].}
	\label{fig4}
\end{figure}

It can be noted that while increasing the 2$\alpha$- and 3$\alpha$-combinations in the event, the manifestation of unstable $^8$Be and HS also increases. The HS identified in the relativistic dissociation of $^{12}$C also appears in the case of $^{16}$O. This result has shown that HS does not reduce to the usual excitation of the $^{12}$C nucleus, but, like $^8$Be, is a more universal object of the nuclear-molecular nature. The most probable confirmation of this assumption can be  observation of the HS in the relativistic fragmentation $^{14}$N $\to$ 3$\alpha$ [13]. Nevertheless, the above observation deserves to be verified for heavier nuclei, where $\alpha$-combinatorics increases rapidly with the mass number.

\section{ALPHA PARTIAL DISSOCIATION OF HEAVY RELATIVISTIC NUCLEI}

\begin{figure}[h]
	\center{\includegraphics[width=0.65\linewidth]{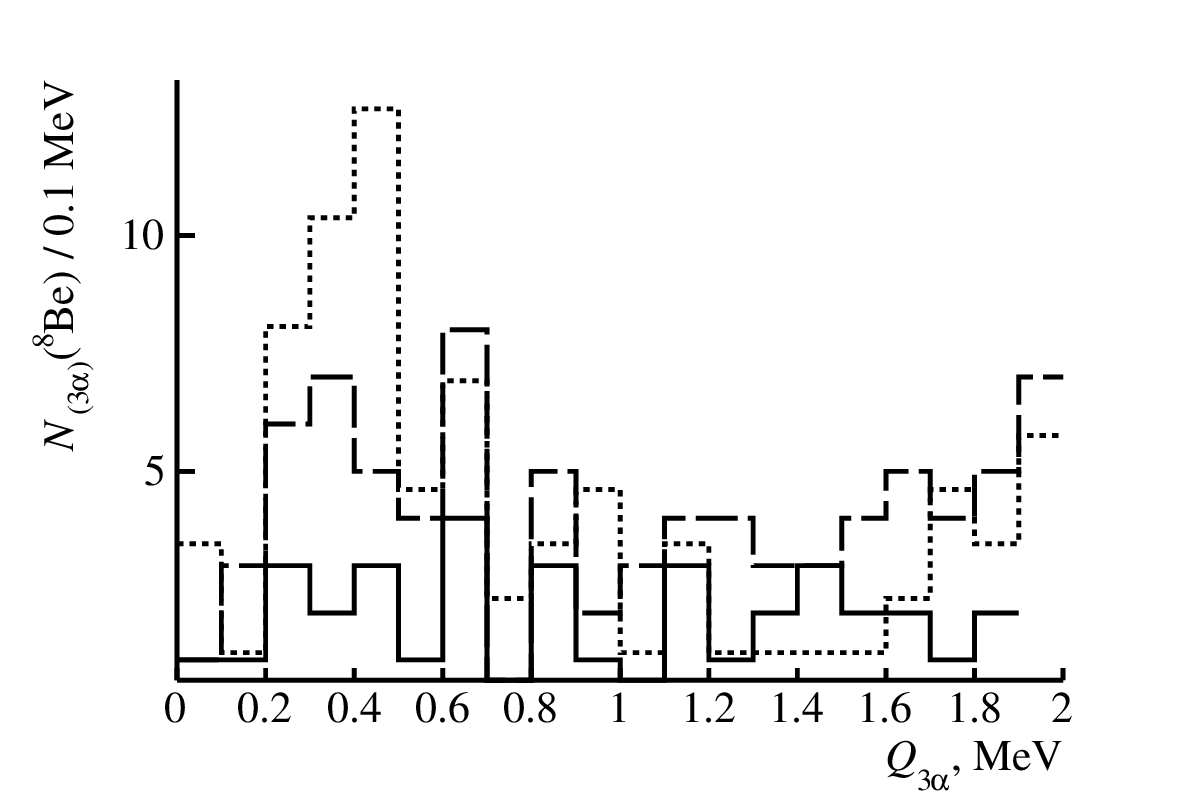}}
	\caption{Distributions of 3$\alpha$ systems with identified $^8$Be $N_{(3\alpha)}$($^8$Be) decays by excitation energy $Q_{3\alpha}$ ($\geq$ 2 MeV) in events of fragmentation of $^{22}$Ne nuclei with an energy of 3.22 GeV/nucleon (solid line) and $^{28}$Si with an energy of 14.6 GeV/nucleon (dashed line). The dots indicate the distribution of $N_{(3\alpha)}$($^8$Be) in the $^{12}$C $\to$ 3$\alpha$ dissociation, normalized to the statistics of $^{22}$Ne and $^{28}$Si [14].}
	\label{fig5}
\end{figure}

Having been tested in the study of light nuclei, a similar search for unstable states was applied to the study of dissociation events of medium and heavy nuclei to identify $^8$Be and HS and search for more complex $n\alpha$BEC states. The analysis has allowed us to trace the contribution of unstable states with a higher multiplicity of the He and H fragments by means of the method of transverse scanning of NTE. Starting with the fragmentation of $^{16}$O nuclei in the energy range from the 3.65 to 200 GeV/nucleon, the analysis has shown a relative increase in the contribution of $^8$Be nuclei with the growing number of relativistic $\alpha$ particles per event [14]. The results of measurements of 4301 interactions of $^{22}$Ne nuclei at the energy of 3.22 $A$ GeV were analyzed in [15]. This data set includes precision measurements of the emission angles of relativistic from 2 to 5$\alpha$ particles, which allowed one to analyse $Q_{(2–5)\alpha}$ variables. It has been found that the probability of identifying the $^8$Be, $^9$B and HS states grows with the multiplicity of $\alpha$ particles in the event [15]. A similar result was obtained by analysing 1093 events of $n\alpha$ fragmentation of $^{28}$Si nuclei with the energy of 14.6 GeV/nucleon in NTE up to 6$\alpha$ particles in the event [14]. Fig. 5 shows the distribution of the events with the identified $^8$Be nuclei by the $Q_{3\alpha}$ variable for the data set on $^{12}$C, $^{22}$N and $^{28}$Si nuclei.

The $n\alpha$ events were studied during transverse scanning of NTE layers longitudinally exposed to the beam of $^{84}$Kr nuclei with the energy of 950 MeV/nucleon [12, 16]. In this analysis, the fragment momentum has been corrected for the ionization losses in NTE and factorized by 0.8 to approximately calculate the drop in the initial momentum value in the interaction [16]. Since it is not sensitive for the selection of $Q_{2\alpha}$($^8$Be) $\geq$ 0.4 MeV, it further allows us to maintain the condition $Q_{3\alpha}$(HS) $<$ 0.7 MeV, focusing on the $Q_{3\alpha}$(HS) peak (Fig. 6). Distributions over $Q_{4\alpha}$ up to 10 MeV have revealed an $\alpha$-quartet at $n\alpha$ = 6 with an isolated value of $Q_{4\alpha}$ = 0.6 MeV, corresponding to both: the $\alpha$HS and 2$^{8}$Be variants [12]. Without contradicting of the above to the decay $^{16}$O(0$^+_6$), this single observation serves as the starting point to further accumulate the statistics on the problem of the 4$\alpha$BEC state.

\begin{figure}[h]
	\center{\includegraphics[width=0.65\linewidth]{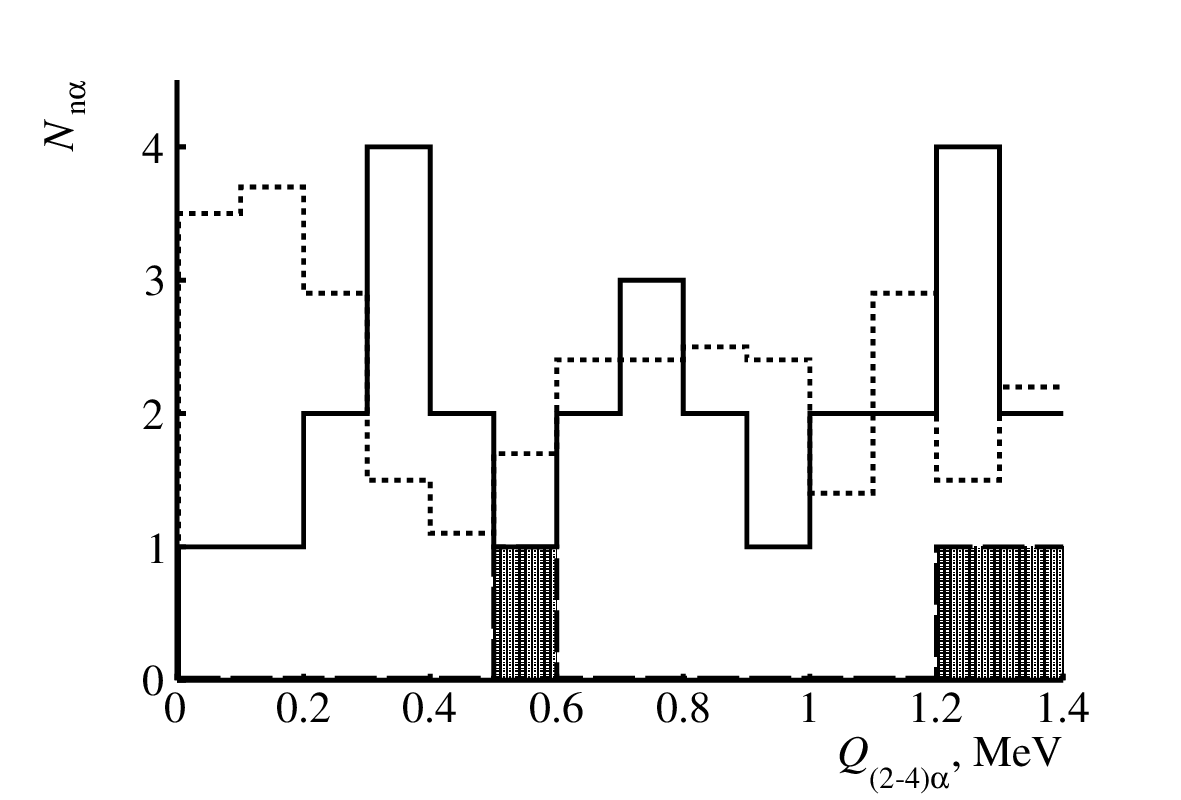}}
	\caption{Distribution in the region of low values of excitation energies of $n\alpha$-systems $Q$: pairs (dots, factor 0.1), triplets (solid line) and quadruples (shaded) $\alpha$-particles formed in the fragmentation of Kr nuclei. [12].}
	\label{fig6}
\end{figure}

\begin{figure}[h]
	\center{\includegraphics[width=0.8\linewidth]{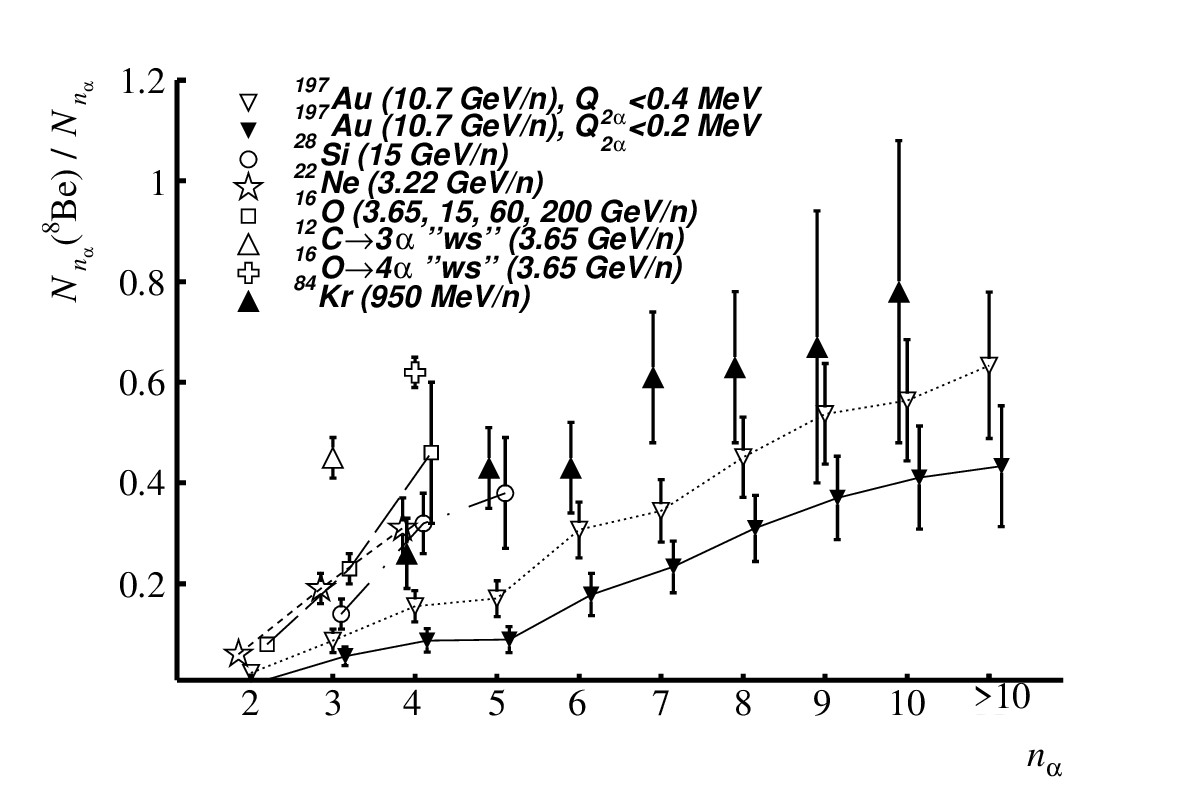}}
	\caption{Dependence of the relative contribution of decays of $^8$Be nuclei $N_{n\alpha}$($^8$Be) to the statistics of $N_{n\alpha}$ events with $\alpha$-particle multiplicity $n\alpha$ in the relativistic fragmentation of C, O, Ne, Si and Au nuclei; ``white'' stars $^{12}$C $\to$ 3$\alpha$ and $^{16}$O $\to$ 4$\alpha$ (WS) are marked; for convenience, the points are slightly shifted around the $n\alpha$ values and connected by lines.}
	\label{fig7}
\end{figure}

The wide coverage of $n\alpha$ has been provided by measuring 1316 inelastic interactions of $^{197}$Au at 10.7 GeV/nucleon [14]. The proportion of the events with $n\alpha$ $>$ 3 among the measured ones was 16\%. Since the complexity of measurements was growing, the selection condition for $Q_{2\alpha}$($^8$Be) has been reduced to $\geq$ 0.4 MeV. It turns out that the ratio of the number of $N_{n\alpha}$($^8$Be) events with at least one identified decay of $^8$Be to their number $N_{n\alpha}$ has demonstrated a strong increase with $n\alpha$ [14,17]. In general, the correlation picture of the dependence of the multiplicity of $\alpha$-particles in one event on the number of events with identified $^8$Be decays (at least one) is shown in Fig. 7.

%\bibliographystyle{unsrt}
%\bibliography{ref}

\end{document}